\documentclass[11pt,twoside]{article}
\usepackage{./asp2014}

\usepackage{floatrow}

\aspSuppressVolSlug
\resetcounters

\begin{document}

\title{The Problem of the High Iron Abundance in Accretion Disks around Black Holes}

\author{J.~A. Garc\'ia,$^{1,2}$ T.~R. Kallman,$^3$ M.~Bautista,$^4$ C.~Mendoza,$^4$ J. Deprince,$^5$ P.~Palmeri,$^5$ and P.~Quinet$^{5,6}$ \\
\affil{$^1$Dept. of Astronomy, California Institute of Technology, Pasadena, CA 91125, USA; \email{javier@caltech.edu}}
\affil{$^2$Remeis Observatory \& ECAP, Universit\"at Erlangen-N\"urnberg, D-96049 Bamberg, Germany}
\affil{$^3$NASA Goddard Space Flight Center, Code 662, Greenbelt, MD 20771, USA; \email{timothy.r.kallman@nasa.gov}}
\affil{$^4$Dept. of Physics, Western Michigan University, Kalamazoo, MI 49008, USA;  \email{manuel.bautista@wmich.edu} \email{claudio.mendozaguardia@wmich.edu}}
\affil{$^5$Physique Atomique et Astrophysique, Universit\'e de Mons, B-7000 Mons, Belgium; \email{jerome.deprince@umons.ac.be} \email{patrick.palmeri@umons.ac.be} \email{pascal.quinet@umons.ac.be}}
\affil{$^6$IPNAS, Universit\'e de Li\`ege, B-4000 Li\`ege, Belgium}
}

\paperauthor{Javier~A.~Garc\'ia}{javier@caltech.edu}{}{California Institute of Technology}{Dept. of Astronomy}{Pasadena}{CA}{91125}{USA}

\paperauthor{Timothy~R.~Kallman}{timothy.r.kallman@nasa.gov}{}{NASA Goddard Space Flight Center}{Laboratory for High Energy Astrophysics}{Greenbelt}{MD}{20771}{USA}

\paperauthor{Claudio~Mendoza}{claudio.mendozaguardia@wmich.edu}{}{Western Michigan University}{Dept. of Physics}{Kalamazoo}{MI}{49008}{USA}

\paperauthor{Patrick~Palmeri}{patrick.palmeri@umons.ac.be}{}{Universit\'e de Mons}{Physique Atomique et Astrophysique}{Mons}{}{B-7000}{Belgium}

\paperauthor{Pascal Quinet}{pascal.quinet@umons.ac.be}{}{Universit\'e de Mons}{Physique Atomique et Astrophysique}{Mons}{}{B-7000}{Belgium}

\paperauthor{Manuel~Bautista}{manuel.bautista@wmich.edu}{}{Western Michigan University}{Dept. of Physics}{Kalamazoo}{MI}{49008}{USA}

\paperauthor{J\'er\^ome~Deprince}{jerome.deprince@umons.ac.be}{}{Universit\'e de Mons}{Physique Atomique et Astrophysique}{Mons}{}{B-7000}{Belgium}


\begin{abstract}
  In most accreting black-hole systems the copious X-rays commonly observed from the inner-most regions are accompanied by a reflection spectrum. The latter is the signature of energetic photons reprocessed by the optically thick material of an accretion disk. Given their abundance and fluorescence yield, the iron K-shell lines are the most prominent features in the X-ray reflected spectrum. Their line profiles can be grossly broadened and skewed by Doppler effects and gravitational redshift. Consequently, modeling the reflection spectrum provides one of the best methods to measure, among other physical quantities, the black-hole spin. At present the accuracy of the spin estimates is called into question because the data fits require very high iron abundances: typically several times the solar value. Concurrently no plausible physical explanation has been proffered for these black-hole systems to be so iron rich. The most likely explanation for the supersolar iron abundances is model shortfall at very high densities ($>10^{18}$~cm$^{-3}$) due to atomic data shortcomings in this regime.  We review the current observational evidence for the iron supersolar abundance in many black-hole systems, and show the effects of high density in state-of-the-art reflection models. We also briefly discuss our current efforts to produce new atomic data for high-density plasmas, which are required to refine the photoionization models.
\end{abstract}

\section{Introduction}\label{intro}

Accreting black holes are readily observable through their intense emissions of high-energy radiation predominantly in the X-ray band. Typically the presence of a compact object is evidenced by its interaction with the surrounding material. This is the case for both stellar-mass black holes in binary systems (black-hole binaries, BHBs) and supermassive black holes in active galactic nuclei (AGN). Importantly, the accretion physics in these two systems---very different in scale---is remarkably similar. The accreted gas trapped by the strong gravitational potential makes its way to the center by spiraling inwards on Keplerian orbits. To preserve angular momentum, the infalling material forms a flat rotating structure known as an {\it accretion disk.} Disk thermal emission is often observed in the ultraviolet (UV) spectrum from AGN \citep[e.g.,][]{kro99}, and in X-rays from galactic black holes \citep[GBHs,][]{don07}. Additionally, an emission nonthermal power-law component is ubiquitous, which arises from much hotter electrons ($kT\sim100$~keV) in an optically thin region referred to as the {\it corona}.

\begin{figure}[h!]
  \centering
  \includegraphics[width=1.0\textwidth]{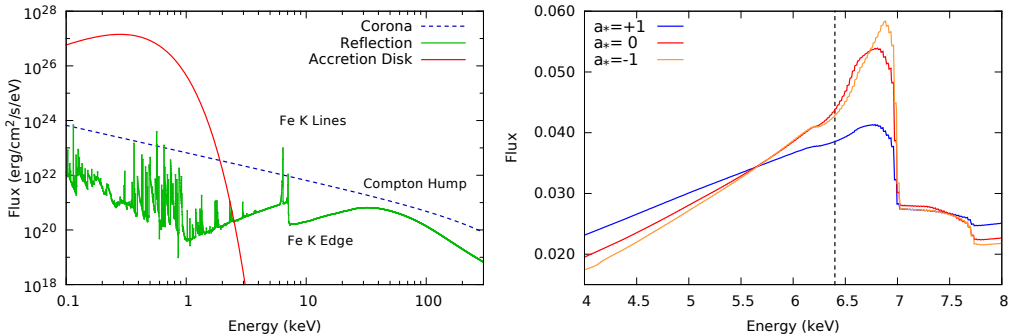}
  \caption{{\it Left}: Three principal components of an accreting BH: (1) the emission power-law coronal component (blue dashed line) that illuminates the accretion disk; (2) the relatively faint and complex reflection component (green) that delivers estimates of disk inclination $i$ and spin $a_*$, the key spectral features (shown here excluding relativistic effects) being the Fe K line/edge and broad Compton hump; and (3) the thermal disk component (red) shown here for $kT \sim 0.1$~keV. In AGN this component can be much colder while in BHBs it can rise up to $kT \sim 1$~keV \citep{rem06}. {\it Right}: Spin effect on the relativistically blurred Fe K line for a disk inclination of $40^{\rm \circ}$. In passing from extreme retrograde spin ($a_*=-1$) to extreme prograde ($a_*=1$), the radius of the ISCO (innermost stable circular orbit) shrinks from $9R_g$ to $1R_g$ (where $R_g = GM/c^2$ is the gravitational radius), gravity builds, and the line broadens dramatically.} \label{fig:xillver}
\end{figure}

The hard coronal radiation illuminates the cold accretion disk and creates the spectral component central to this proposal---the {\it reflection} spectrum---which is a forest of fluorescent lines, edges, and related features (Fig.~\ref{fig:xillver}, left). This component leaves the disk carrying information on the physical composition and condition of the matter in the strong fields near the black hole. The most prominent feature is the fluorescent Fe K complex of emission lines at 6.4--6.9~keV. The line profiles are grossly distorted in the strong-gravity regime by Doppler effects, light bending, and gravitational redshift (Fig.~\ref{fig:xillver}, right). By modeling the shape of the Fe K profile (as well as the entire reflection spectrum), much can be deduced about the matter near the black hole and the black hole itself, including its rotation rate ({\it spin}).

However, the accuracy of these estimates depends strongly on the microphysics; in particular, the iron abundance plays a significant role in the correct modeling of the spectral features. In fact, some of the spin estimates have been questioned due to the very large abundances required to fit the observational data. In this paper we review the current state of these predictions, and discuss the impact of high-density plasma effects in the calculation of the reflected spectrum. We also briefly discuss the shortage of atomic data appropriate to model such environments and our ongoing efforts to improve current photoionization models.

\section{Theoretical and Observational Methods}

\subsection{Reflection Spectroscopy}\label{sec:refl}

X-ray reflection spectroscopy is arguably the most effective approach currently available for probing the effects of strong gravity near an event horizon. Reflection models are the basis of the Fe-line method for measuring spin, a technique widely used for both supermassive black holes \citep[e.g.,][]{pat11,wal13,rey14} and stellar black holes \citep[e.g.,][]{mil07,rei08,mcc14,gar15b}. Furthermore, it is used to infer many important properties of the accretion process by characterizing the gas fueling the black hole \citep{gar15a}. Theoretical models of X-ray reflection have been undergoing active development over the past three decades \citep[see][for a  review]{fab10}. Currently the most updated and advanced models are {\sc xillver} \citep{gar10,gar13a} and its relativistic counterpart {\sc relxill} \citep{dau13,gar14a}. They have proven to be effective in probing the relativistic effects due to the strong gravitational field near black holes.

\begin{figure}[t!]
  \centering
  \includegraphics[width=0.72\textwidth]{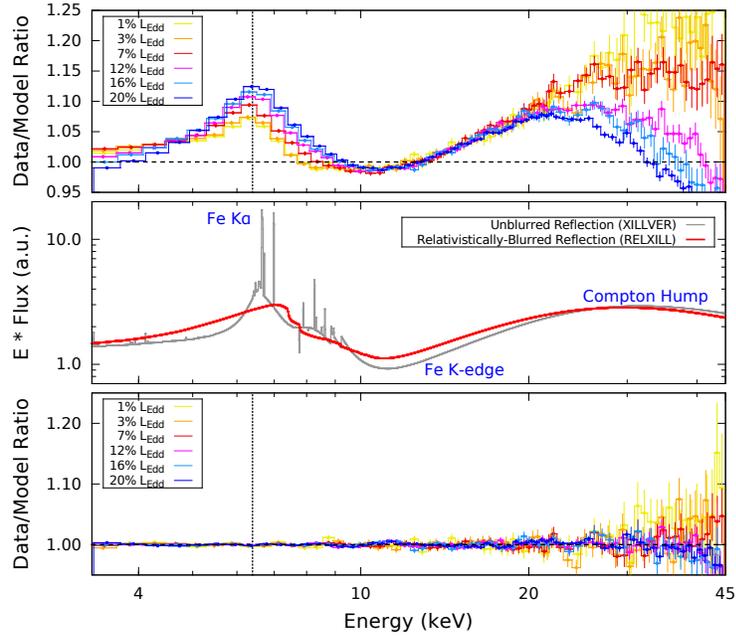}
  \caption{Spectral analysis of {\it RXTE} PCA observations for GX~339--4 at six different accretion rates in the hard state \citep{gar15b}. {\it Top}: Ratio to a power law model (the residual features are fingerprints of unmodeled reflection). {\it Middle}: Reflected spectra predicted by the {\sc xillver} model (gray) and its relativistically blurred counterpart {\sc relxill} (red). The main reflection signatures are labeled. {\it Bottom}: Spectrum ratio after including reflection. [Adapted from Fig.~6 of \citet{gar15b}.]} \label{fig:gx339}
\end{figure}

In Fig.~\ref{fig:gx339} we illustrate an application of these calculations by modeling high-signal {\it RXTE} spectra of the stellar black hole system GX~339--4 \citep{gar15b}. Six spectra are shown covering a range in luminosities from 17\% to 1.6\% of the Eddington limit ($L_\mathrm{Edd}\sim 10^{39}$~erg~s$^{-1}$). These constitute one of the highest signal-to-noise reflection measurement to date.  Our model provides an excellent fit while delivering precise constraints on the black hole's spin
($a_*=0.95^{+0.03}_{-0.05}$), disk inclination ($i=48\pm1$~deg), and iron abundance ($A_\mathrm{Fe}=5\pm1$ solar).

\begin{figure}[h!]
  \centering
  \includegraphics[width=0.72\textwidth]{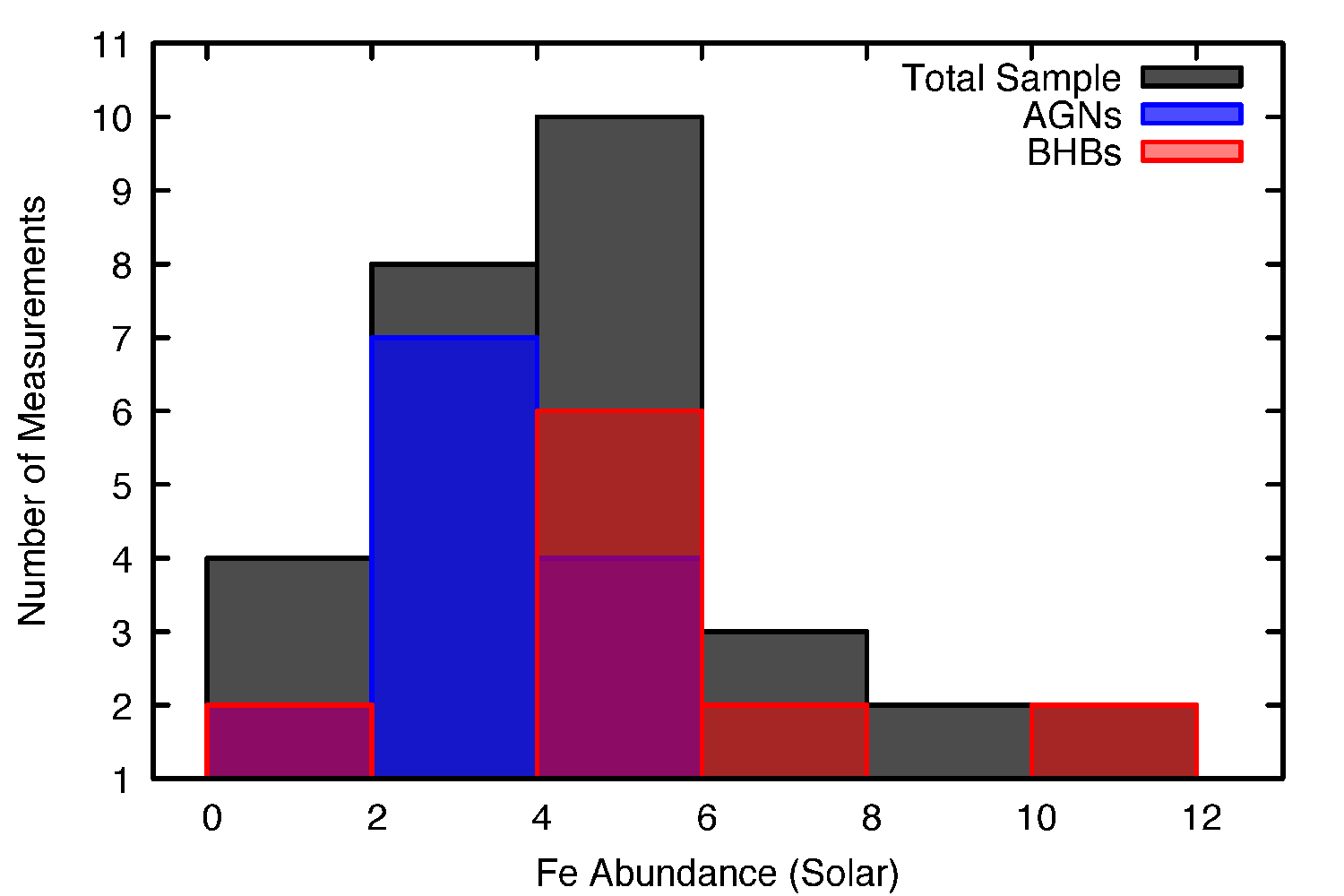}
  \caption{Histogram of iron abundance determinations using reflection spectroscopy for both 13 AGN and 9 BHBs. Values a few times the solar standard are routinely found.}\label{fig:rin_ironabund}
\end{figure}

This supersolar prediction for the Fe abundance is not unique to GX~339--4. In fact a similar result is found in other  stellar-mass black hole binaries such as V404~Cyg \citep{wal17} and Cyg~X--1 \citep{par15}. Even more interesting is the similar trend found in AGN as well, the most dramatic case being the Seyfert galaxy 1H0707--495 for which Fe is overabundant by a factor of 10--20 \citep{fab09,dau12}. Fig.~\ref{fig:rin_ironabund} shows a compilation of the iron abundance reported in the recent literature by implementing our reflection models for 13 AGN and 9 BHBs. Despite the relatively small sample, the trend is clear in both cases for abundances a few times over the solar value. No satisfactory explanation has yet been proposed, which motivates the examination of the line-emission microphysics in the current reflection models.

\subsection{X-Ray Reflection at High Density}\label{sec:line}

Most reflection models currently available consider spectra calculated at relatively low densities ($n_e\sim10^{15}$~cm$^{-3}$). However, predictions based on the standard $\alpha$-disk model \citep{sak73} and more sophisticated 3D magneto-hydrodynamic (MHD) simulations \citep[e.g.,][]{nob10,sch13} suggest densities in black-hole accretion disks orders of magnitude larger.  Moreover, observations of several AGN reveal extreme illumination of the inner regions of the accretion disk
\citep[e.g.,][]{par14,kar15b}, which indicate coronae very close to the inner disk. The observed luminosities and ionization parameters imply densities much larger than the values typically assumed for AGN disks.

\begin{figure}[h!]
  \centering
  \includegraphics[width=\textwidth]{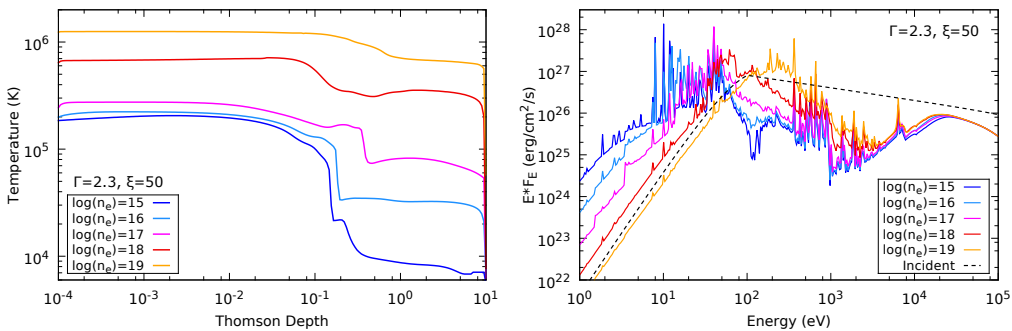}
  \caption{Reflection calculations for various densities using parameters appropriate to AGN. {\it Left}: Temperature profiles in the vertical direction of the illuminated atmosphere. Each curve corresponds to a different value of density while the ionization parameter is held constant. {\it Right}: Angle-averaged reflected spectra. As indicated, solid curves correspond to different density values. The dashed curve is the incident power-law spectrum that illuminates the disk and is the same for all models. [Reproduced from Figs. 2 and 4 of \citet{gar16b} by permission of Oxford University Press.]}\label{fig:lowXi}
\end{figure}

We have recently studied the effects of higher-density plasmas in computing X-ray reflected spectra using our {\sc xillver} model. By performing calculations over a range of densities, we demonstrate that, at sufficiently high densities ($n_e > 10^{17}$~cm$^{-3}$), ionization effects result in a marked increase in atmospheric temperature \citep{gar16b}. The main effects of increasing density are the suppression of line cooling near the surface and the increase of free--free heating in the deeper regions, which result in an overall increase of the gas temperature (Fig.~\ref{fig:lowXi}, left). Consequently, the ionization balance is also affected typically increasing the gas ionization state as the density increases (Fig.~\ref{fig:lowXi}, right).  The most obvious effect in the reflected spectrum is the modification of the thermal emission at soft energies that follows a Rayleigh--Jeans law. Since the temperature is larger at high-densities, the emission peak moves to higher energies. Additionally, the larger temperature also leads to a higher ionization that affects line emission and continuum photoelectric absorption. In general the reflected spectrum at energies below ${\sim}2$~keV shows a significant flux excess when compared to lower density spectra.

While the mentioned effects do not have an obvious and direct effect on the Fe K-line profile, the dramatic change of the reflected continuum at low energies may have an indirect impact on the Fe abundances derived from spectral fitting.  This has been recently demonstrated for the BHB Cyg~X--1. Analysis of {\it NuSTAR} and {\it Suzaku} observations of this source revealed the presence of a strong soft-excess component at low energies. Fits with standard reflection models returned very high and unphysical Fe abundances (above 10 times solar). \cite{tom18} showed that, by implementing high-density reflection models, the change in the reflected continuum provides a different fit that mitigates the need for the extremely large abundances.
This results suggest that an improper modeling of the reflected continuum at soft energies could also be responsible for the supersolar abundances observed in some systems.

\subsection{Reflection at Even Higher Densities}

To date our reflection calculations have been extended to densities up to $n_e\sim10^{19}$\,cm$^{-3}$. Higher densities are not accessible due to limitations in the current atomic data. In particular, recombination rates and other atomic quantities have not been tabulated for densities outside this range. Importantly, screening effects from neighboring atoms at densities above $n_e\sim10^{19}$~cm$^{-3}$ become relevant.

High plasma densities in particular can be expected to affect all relevant atomic processes, either by truncating the bound levels with high principal quantum number (continuum lowering), increasing the importance of collisional processes, or changing the effective nuclear charge and thereby the atomic structure and associated rates. Some of these are competing effects: stimulated recombination is enhanced, but both recombination into high-$n$ levels and dielectronic recombination are suppressed. For Fe K-vacancy levels and their respective decays at such densities ($n_e\lesssim 10^{21}$\,cm$^{-3}$), the changes in $A$-values in \ion{Fe}{xxv} and \ion{Fe}{xxiv} were found to be less than 1\% \citep{dep17} but could be much higher at densities $n_e > 10^{21}$\,cm$^{-3}$ and in lower charge states. The charge-state distributions will also be very different, and the lines from high-$n$ levels will be largely quenched at high density with exception of transitions from H- and He-like ions.

We are presently involved in a systematic project to compute atomic parameters taking into account the plasma effects arising at high densities. In particular, \citet{dep18a} have estimated the effects of the plasma environment on the atomic parameters associated with K-vacancy states in He- and Li-like oxygen ions using multiconfiguration Dirac--Fock and  Breit--Pauli approaches. These computations have been carried out assuming a time-averaged Debye--H\"uckel potential for both the electron--nucleus and electron--electron interactions. Plasma effects on the ionization potential, excitation thresholds, transition wavelengths, and radiative emission rates are reported therein for \ion{O}{vi} and \ion{O}{vii}. The same methodology is also being applied to compute the plasma effects on the atomic structure and radiative and Auger decay rates of the K-lines of \ion{Fe}{xvii} -- \ion{Fe}{xxv} \citep{dep18b}.

\section{Final Remarks}

The problematic large iron abundances required in reflection models to fit the X-ray spectra from accreting black holes is still an open question. Given that this strong requirement is common to both AGN and BHBs, it is unlikely that supersolar
abundances are realistic since metal enrichment mechanisms in these two types of systems are expected to be very different. We have discussed how high-density plasma effects can help to mitigate this problem. So far we have found the effects of
high density in the determination of the Fe abundance via reflection spectroscopy are (at least) two-fold: (1) indirect, by affecting the shape of the reflected continuum at soft energies, where reflection from a high-density atmosphere will produce a flux excess due to free--free heating; and (2) direct, by affecting the atomic parameters that control line emission and photoelectric absorption. In the second scenario, continuum lowering is likely to be one of the most significant factors in controlling the iron abundance. Our efforts are currently in progress in the investigation of this important question.

\acknowledgements
This project is currently being funded by the NASA Astrophysics Research and Analysis Program, grant 80NSSC17K0345. JAG acknowledges support from the Alexander von Humboldt foundation. JD is a Research Fellow of the Belgian Fund for Research Training in Industry and Agriculture (FRIA). PP and PQ are, respectively, Research Associate and Research Director of the Belgian Fund for Scientific Research (FRS-FNRS).



\end{document}